\documentclass{epl}

\newcommand{\ie}{{\it i.e.},}
\newcommand{\eg}{{\it e.g.},}
\newcommand{\gpd}{\ensuremath{\varphi}}
\newcommand{\kb}{\ensuremath{k}_{\rm B}}
\newcommand{\e}{{\rm e}}
\newcommand{\Ss}{{\mathcal S}}
\newcommand{\F}{{\mathcal F}}
\newcommand{\eps}{{\epsilon}}

\title{\vspace{-8mm}{\it\small Comment}\\ Relation between the thermodynamic
Casimir effect in Bose-gas slabs and critical Casimir forces%
}
\author{A.~Gambassi\inst{1,2}\thanks{E-mail: \email{gambassi@mf.mpg.de}} \and S.~Dietrich\inst{1,2}}

\institute{                    
  \inst{1} Max-Planck Institut f\"ur Metallforschung -
  Heisenbergstr. 3, D-70569 Stuttgart, Germany \\
  \inst{2} Institut f\"ur Theoretische und Angewandte Physik,
  Universit\"at Stuttgart - Pfaffenwaldring 57, D-70569 Stuttgart, Germany
}

\begin{document}

\maketitle

%\begin{spacing}{2.5}

\begin{abstract}
In a recent letter, Martin and Zagrebnov [{\it Europhys. Lett.,} {\bf
73} (2006) 1] discussed the {\it thermodynamic} Casimir effect for the
ideal Bose gas confined in a thin film. 
We point out that their findings can be expressed in terms of 
previous general results for the 
Casimir effect induced by confined {\it critical fluctuations}. This
highlights the links between the Casimir effect in the contexts of
critical phenomena and Bose-Einstein condensation. 
\end{abstract}

The ideal Bose gas undergoes a phase
transition in the grand canonical ensemble if the
chemical potential $\mu \le 0$ equals its critical value $\mu_c =
0$ at which Bose-Einstein condensation takes place. This
transition is accompained by density fluctuations with an increasing 
correlation length 
$\xi_+(\mu\rightarrow \mu_c) \sim (\mu_c - \mu)^{-\nu}$ ($\nu =
\frac{1}{2}$ for the ideal gas).

In ref.~\cite{paper} the authors consider the limit of large film
thicknesses $d$ and
{\it fixed} $\mu$ of the grand canonical
potential per unit (transverse) area $\gpd_d(T,\mu)$ of an ideal Bose
gas in spatial dimension $D=3$. 
For $\mu \neq \mu_c$ this implies that eventually $d/\xi_+ \gg 1$ and
therefore the confining boundaries are subject to a vanishing Casimir
force resulting from correlated fluctuations. 
On the other hand, at the critical point $\mu=\mu_c$, $d/\xi_+
\rightarrow 0$ and the
fluctuations become long-ranged, giving rise to a Casimir force per
unit area $F(d,T,\mu=\mu_c) = 2 \kb T \Delta/d^3 + \ldots$, characterized by a
{\it universal} amplitude $\Delta$%
%
%%%%%%%%%%%%%%%%%%%%%%%%%%%%%%
\footnote{%%
In passing we note
that the amplitudes 
$\Delta(T)$ introduced in eqs.~(3), (12), and (13) in
ref.~\cite{paper} are not {\it
universal}, unless they are properly normalized as $\Delta \equiv \Delta(T)/\kb
T$.}%
%%%%%%%%%%%%%%%%%%%%%%%%%%%%
. The cross over between these two regimes, which has not been
investigated in ref.~\cite{paper}, is determined by 
$\delta \gpd_d(T,\mu) \equiv \gpd_d(T,\mu)
- d\, \varphi_{bulk}(T,\mu) - \varphi_{surf}(T,\mu)$ (see eqs.~(2) and (9)
in ref.~\cite{paper}) for large $d$ and {\it fixed} ratio 
$d/\xi_+ \sim d (-\mu)^\frac{1}{2}$. Here we consider the case of
Dirichlet boundary conditions (BCs), although the extention to other cases
is straightforward. According to eq.~(17) in ref.~\cite{paper} one has:
\begin{equation}
- \frac{(2 \pi)^\frac{3}{2}}{2} \beta d^2 \delta \gpd_d(T,\mu) = 
\left(\frac{d}{\lambda}\right)^3
\sum_{n,r = 1}^\infty \frac{\e^{\beta \mu r}}{r^{5/2}} \e^{- 2 (n
d/\lambda)^2/r} = \sum_{n=1}^\infty \left(\frac{\lambda}{d}\right)^2
\sum_{r=1}^\infty \Phi_n((\lambda/d)^2 r, u) ,
\end{equation} 
where $\beta \equiv 1/(\kb T)$, $\lambda \equiv \hbar\sqrt{\beta/m}$ is the
thermal wavelength, $\Phi_n(s,u) \equiv s^{-\frac{5}{2}}\e^{- u^2 s/2 - 2
n^2/s}$, and $u\equiv (-2\beta \mu)^\frac{1}{2} d/\lambda \sim d/\xi_+$. 
For $d/\lambda \gg 1$, $(\lambda/d)^2 \sum_{r}
\Phi_n((\lambda/d)^2 r, u) \mapsto \int_0^\infty \upd s\,
\Phi_n(s,u)$,
with corrections decaying faster than any
power of $\lambda/d$, leading to~\cite{GR}
\begin{equation}
\beta d^2 \delta \gpd_d(T,\mu) = -\frac{1}{8\pi} \sum_{n=1}^\infty
\frac{1 + 2 u\, n}{n^3} \e^{-2 u\, n} \equiv \Theta(u) ,
\label{conn}
\end{equation}
so that $F(d,T,\mu) = - \partial
\delta\gpd_d(T,\mu)/\partial d = \kb T [2\Theta(u) - u\Theta'(u)]/d^3$.
$\Theta(u)$ is a continuous, negative, and monotonicly 
increasing function of $u$ providing the 
interpolation between the two cases discussed in
ref.~\cite{paper}: 
At the transition point $u=0$ one recovers eq.~(13) therein, %in ref.~\cite{paper},
\ie\  
$\Delta \equiv \Theta(0) = -\zeta(3)/(8\pi)$. Due to $\Theta(u\gg
1) \sim \e^{-2 u}$  for $d\gg \lambda$
and fixed $\mu$ we recover eq.~(21): 
$|\delta \gpd_d(T,\mu)| \le 
O(\e^{-\sqrt{-8\beta\mu}d/\lambda})$. It is easy to verify that $\Theta(u) =
\Theta^{(1)}_{+O,O}(u)$ in $D=3$ and with $N=2$, where 
$\Theta^{(1)}_{+O,O}$ is given by eq.~(6.6)%
%%%%%%%%%%%%%%%%%%%%%%%%%%%%%%%%%%
\footnote{%
The correct lower integration limit is $x=1$ in accordance with eq.~(6.5).}
%%%%%%%%%%%%%%%%%%%%%%%%%%%%%%%%%%
in ref.~\cite{KD-92}. $\Theta^{(1)}_{+O,O}$ is 
the Gaussian, $(1)$, {\it universal} scaling function above the bulk
critical temperature, $+$, for
the finite-size contribution $\beta d^{D-1}\delta \F_d(T)$ to the free
energy $\F_d(T) = d \F_{bulk}(T) + \F_{surf}(T) + \delta \F_d(T)$ 
per unit (transverse) area of systems
the critical properties of which are captured by
the $O(N)$ symmetric Landau-Ginzburg (LG) Hamiltonian~\cite{ZJ-book},
confined in a
$d\times\infty^{D-1}$ slab with Dirichlet 
BCs, $O,O$.
This unnoticed connection between the results of refs.~\cite{KD-92}
and~\cite{paper} holds also for periodic and Neumann 
BCs considered in ref.~\cite{paper}.
It is rooted in the fact that
the grand canonical partition function for a weakly interacting 
Bose gas in $D$ dimensions 
can be expressed, close to the transition point ($D>2$), as a
functional integral with weight $\e^{-\Ss[\phi]}$ (see, \eg\
refs.~\cite{BBZJ-00,ZJ-book})  where $\phi(x)$ is a two-component
real field, $\Ss$ the $O(2)$ LG Hamiltonian
\begin{equation}
\Ss[\phi] = \int \upd^Dx\, \left\{ \frac{1}{2} [\nabla \phi(x)]^2 +
\frac{1}{2} r \phi^2(x) + \frac{g}{4!} [\phi^2(x)]^2\right\} ,
\end{equation}
$r = - 2 m \mu/\hbar^2$, and $g = 48 \pi a \hbar^4/\lambda^2$ where $a$ is
the scattering length. 
For the ideal gas $g=0$, and $\Ss$ reduces to the
so-called Gaussian model (defined only for $r>0$), 
characterized by $\xi_+ = r^{-\frac{1}{2}}$. $\Ss$ is a particular case of the more general
$O(N)$-symmetric LG Hamiltonian considered in ref.~\cite{KD-92}, where 
the {\it universal} scaling
functions $\Theta(y_+)$ corresponding to different BCs
({\it surface universality classes}) imposed on $\phi(x)$
have been determined analytically within the field-theoretical
$\eps$-expansioon ($\eps = 4 - D$), as functions of the scaling
variable $y_+ \equiv d/\xi_+$. For the ideal Bose gas $y_+ = d
r^\frac{1}{2} = u$, as in eq.~(\ref{conn}). 
Thus, the general results of ref.~\cite{KD-92} predict also the
Casimir amplitude $\Delta_{\rm int}$ in a {\it weakly interacting}
Bose gas (as well as for the superfluid transition, belonging to the
same $XY$ universality class $N=2$). In $D=3$ and for Dirichlet 
BCs $\Delta_{\rm int} \simeq  -0.022$ compared to the ideal gas case
$\Delta \simeq -0.048$.

%%%%%%%%%%%%%%%%%%%%%%
%%%%%%%%%%%%%%%%%%%%%%

%\end{spacing}

\end{document}